# Title: Room temperature Silicon detector for IR range coated with Ag$_2$S Quantum Dots

*Ivan Tretyakov*[1*], *Alexander Shurakov*[1,2], *Alexey Perepelitsa*[1,4], *Natalya Kaurova*[1], *Sergey Svyatodukh*[1], *Tatyana Zilberley*[5], *Sergey Ryabchun*[1,3], *Mikhail Smirnov*[4], *Oleg Ovchinnikov*[4] *and Gregory Goltsman*[1,2,3]

[1] Moscow State University of Education, Moscow 119435, Russia.

[2] Zavoisky Physical-Technical Institute of the Russian Academy of Sciences, Kazan 420029, Russia

[3] National Research University Higher School of Economics, Moscow 101000, Russia.

[4] Voronezh State University, Voronezh 394018, Russia.

[5] Moscow Institute of Physics and Technology (State University), Dolgoprudny 141701, Russia.

**E-mail:** ivantretykov@mail.ru

**Keywords:** infrared range, silicon, quantum dot, surface states, room-temperature detector.

**Abstract:** For decades silicon has been the chief technological semiconducting material of modern microelectronics and has had a strong influence on all aspects of society. Applications of Si-based optoelectronic devices are limited to the visible and near infrared ranges. For photons with energy less than 1.12 eV silicon is almost transparent. The expansion of the Si absorption to shorter wavelengths of the infrared range is of considerable interest to optoelectronic applications. By creating impurity states in Si it is possible to cause sub-band gap photon absorption. Here, we present an elegant and effective technology of extending the photoresponse of towards the IR range. Our approach is based on the use of Ag$_2$S quantum dots (QDs) planted on the surface of Si to create impurity states in Si band gap. The specific sensitivity of the room temperature zero-bias Si_Ag$_2$S detector is $10^{11}$ cm$\sqrt{Hz}W^{-1}$ at 1.55μm. Given the variety of available QDs and the ease of extending the photoresponse of Si towards the IR range, our findings open a path towards the future study and development of Si detectors for technological applications. The current research at the interface of physics and chemistry is also of fundamental importance to the development of Si optoelectronics.

The detection of low power radiation of the IR range presents great challenges to Si-based optoelectronics. The development of effective room-temperature photoconductors for the IR compatible with the current silicon technology is in great demand. In particular, megapixel digital imaging based on complementary-metal-oxide-semiconductor (CMOS) technology in the IR range has a high potential for crucial technological applications such as night vision systems, spectroscopy, medical diagnosis, environmental monitoring, on-chip optical data processing and astronomy.[1-6]

At room temperature Si has an indirect band gap ΔE of 1.12 eV, for this reason Si is transparent for IR radiation from a wavelength of 1.1 μm. This significantly limits the range of applications of current Si photoconductive devices. Various methods have been proposed to extend the photo-response of Si towards longer wavelengths of the IR range. One of them uses "heavy" doping of Si with chalcogen atoms Se and transition metals (Au, Ag, Ti, etc.).[7-11] Such hyperdoped materials offer an absorption coefficient increased by a factor of 2.5 in the IR range. Another approach is to integrate non-silicon electro-optical materials, such as direct-band gap III–V compound semiconductors (GaAs, InAs) with excellent IR photo-responce. Until recently it has been understood that this way is strongly limited by the lattice mismatch between Si and these materials [12]. However, significant progress has recently been made in the manufacture and study of high-quality structures based on materials from groups III-V grown on a Si substrate [13-15] 2D materials do not suffer from this limitation and can be transferred to any substrate and used for IR and THz detection.[16-19] CVD-grown graphene monolayer coated with PbS QDs was monolithically integrated with silicon-integrated circuits based on CMOS for digital imaging in the IR range.[20]

In this paper, we report on the significant room-temperature photo-response of a Si photo-detector in the near-IR (NIR) and short-wave IR (SWIR) The synsebilization is possible by doping the Si surface by individual semiconducting $Ag_2S$ QDs. This results in the formation of impurity states in the band gap of Si, which leads to the significant enhancement of the sub-band gap photo-response. Additionally, it enables room-temperature operation of our detector, since $Ag_2S$ QDs introduce relatively high acceptor levels in the Si band gap, which mitigate thermal carrier generation.

Semiconducting QDs are nanocrystals with a size of the order of the Bohr radius of the Wannier-Mott exciton in the corresponding material.[21] Using QDs of different composition allows adjusting the spectral range of the detector depending on the absorption region of the QD. [22-25] Interest in using QDs is also due to the ease of the adjustment of the optical properties of the QD by changing its size. [23-25] $Ag_2S$ has a relatively narrow band gap of 1.0 eV in bulk and a high concentration of trap states, the presence of which is associated with QD nonstoichiometry.[29-32] The aqueous solutions of $AgNO_3$ and $Na_2S$ were used as the initial reagents aqueous synthesis of colloidal $Ag_2S$ QDs was carried out under conditions similar to described in Ref. 32 and Ref. 33. In order to remove the reaction by-products the acetone was used in aqueous colloidal Ag2S QDs solution followed by centrifuging the solution. Thioglycolic acid molecules were used as stabilizers.

Initial Si structures were fabricated using standard methods of laser lithography, thermal metal deposition and a lift-off process based on undoped high-resistance Si ($\rho > 3$ KΩ*cm). Using laser lithography, the width of the Ti/Au contacts W and the distance between them L were defined, thereby determining the geometric dimensions of the original Si structure, W and L were equal to 10 μm. The contacts to the Si structure were formed by sequential thermal deposition of Ti and Au. **Figure 1**a shows an optical image and design of the investigated devices. During the fabrication process of Si_$Ag_2S$ devices, the quality and purity of the Si surface between the Ti/Au contacts is important. Before depositing colloidal $Ag_2S$ from the solution on top of Si, its surface was cleaned by ion etching in an Ar and $O_2$ atmosphere and liquid etching with hydrofluoric acid. The cleaning processes made it possible to remove of all impurities and natural oxides on the Si surface. $Ag_2S$ QDs from a colloidal solution were centrifuged onto a prepared Si surface between the Ti/Au contacts with the subsequent evaporation of the solvent. **Figure 1**b shows the Si surface between the Ti/Au contacts obtained by SEM after deposition of $Ag_2S$ QDs. Structural properties of the devices were obtained using microscopy and X-ray diffraction analysis. The analysis of the SEM images shows that QDs are formed with average sizes of 2.5 ± 0.5 nm. High resolution SEM images show the diffraction on the (012) atomic plane of monoclinic $Ag_2S$ lattice. This fact indicates the formation of crystalline $Ag_2S$ nanoparticles. The X-ray diffraction data also confirm the formation of $Ag_2S$ nanocrystals with a monoclinic lattice (space group P 21/c). X-ray reflections were broadened due to the size effect. As can be seen from **Figure 1**b, $Ag_2S$ QDs located on the Si surface don't interact with each other.

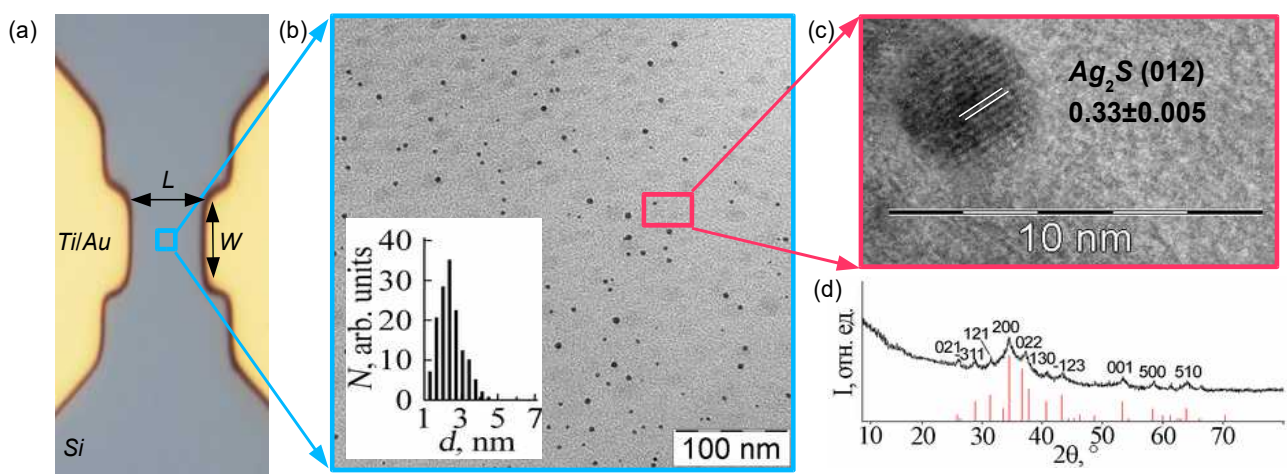

**Figure 1.** a) An optical image of the devices; IR radiation was focused on Si surface between Ti/Au. b) A SEM image of the Si surface after the deposition of $Ag_2S$ QDs; the QDs were formed with average sizes of 2.5 ± 0.5 nm and didn't come into contact with each other. c) A hight-resolution SEM image of $Ag_2S$ QDs on top of Si. The diffraction on the (012) atomic plane of the monoclinic Ag2S lattice indicates the formation of crystalline $Ag_2S$ nanoparticles; d) X-ray diffraction data for $Ag_2S$ nanocrystals with a monoclinic lattice.

The extension of the spectral response and sensitivity for $Ag_2S$_Si devices in the NIR and SWIR ranges were studied experimentally. The devices were shielded from the background radiation and in the absence of IR radiation had a resistance R of 20 MΩ. To couple the devices with IR radiation we used a hyper-hemispherical Si lens. The IR radiation incident on the lens was collected into an Airy spot on the Si surface between the Ti/Au contacts. The studies of the spectral response $S_v(\lambda)$ of our devices were performed in the range of 1 - 2 μm, at 300K and zero bias device mode. **Figure 2** presents $S_v(\lambda)$ for Si and Si_$Ag_2S$ devices. The $S_v(\lambda)$ dependence for Si device is given for bright estimation of the IR band extension for the Si_$Ag_2S$ device. The $S_v(\lambda)$ dependence of the Si device in **Figure 2** has a maximum at 1.1 μm and drops rapidly already at 1.25 μm. At same time, $S_v(\lambda)$ for the Si_$Ag_2S$ device demonstrates a monotonic fall down to 2 μm. The cut-off wavelength of the internal photovoltaic effect for bulk $Ag_2S$ is around 1.25 μm. Considering this fact, the strong response of the Si_$Ag_2S$ device above 1.25 μm can be explained by the formation of "surface states" caused by the $Ag_2S$ QDs coating on the Si surface. In the Si_$Ag_2S$ device, when an IR photon is absorbed by an electron in the valence band of Si, the generated carrier transits at sub-band states $Ag_2S$. This process is detected as a voltage change between the Ti/Au device contacts. For a quantitative evaluation of the Si sensitization effect, it is convenient to bring the ratio of the signal for the Si_$Ag_2S$ structure to the signal for the Si structure taken at a given λ. In our experiment, this ratio was more than 40 at 1.45 μm. Thereby the deposition of $Ag_2S$ QDs on the surface of the Si leads to the formation of devices with a wider absorption band than that of the initial components separately.

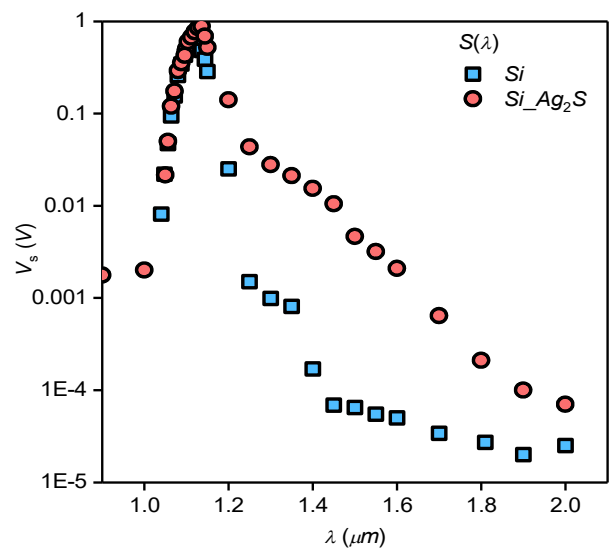

**Figure 2.** Spectral response $S_v(\lambda)$ of Si and Si_Ag$_2$S devices. The curves reflect the effect of Si sensitization caused by the absorption of a sub-band gap IR photon. The electron absorbed IR photon from valence band of Si transits at sub-band surface state Ag$_2$S. This process affects the space charge region near the Ti/Au contacts.

In order to clarify this phenomenon, a current-voltage (IV) curve of the Si_Ag$_2$S device was investigated experimentally. **Figure 3**a shows experimental IV curves of the Si_Ag$_2$S device illuminated by IR light of various power levels at 1.55 μm. As can be seen, at a low bias the IV curves of the device are not symmetric about zero. Referring to **Figure 3**b, the Si_Ag$_2$S device utilizes two metal/semiconductor (MS) Ti/Si junctions. In the ideal case described by the Schottky-Mott theory, each such a junction is characterized by the appearance of a barrier in the vicinity of the MS interface caused by the difference of the metal's work function and the semiconductor's electron affinity. Given that Si possesses the interface behavior parameter of a small value, a space charge region and a built-in potential $\varphi_{bi}$ are formed at the interface which noticeably affects the transport of carriers through the junction near the Ti/Au contacts.[34]

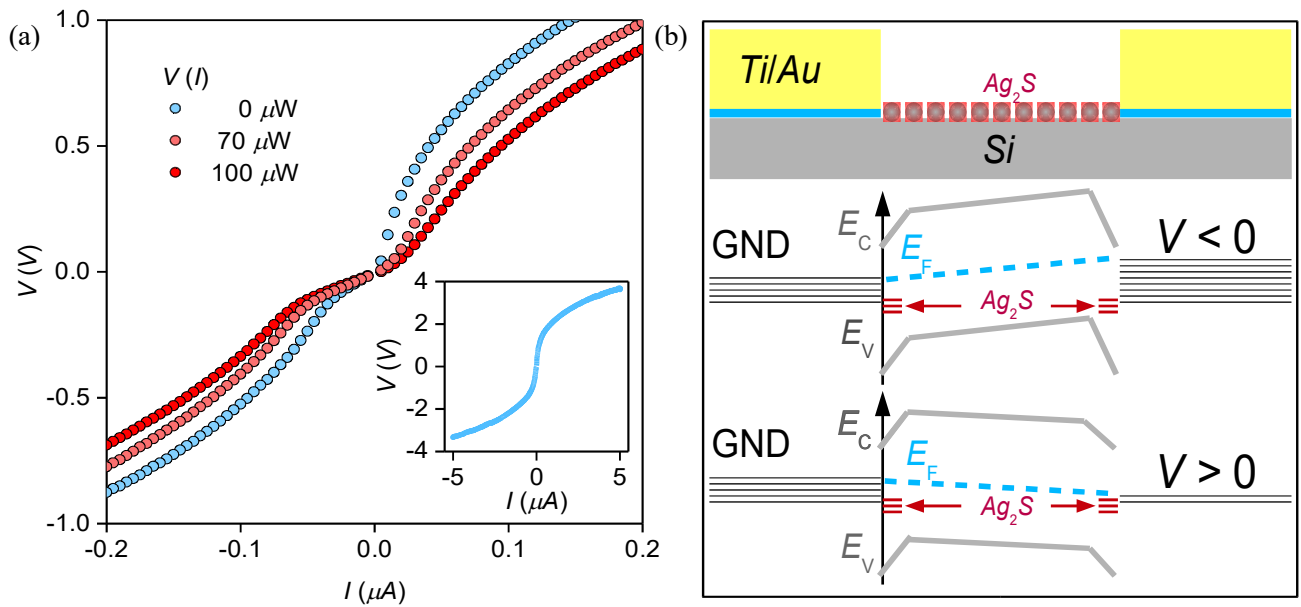

**Figure 3.** a) IV curves of a Si_Ag$_2$S device under different IR power levels at 1.55 μm. The left insert presents an IV curve with no IR power for a wide bias range. The behavior of the IV curve suggests the "hole" conductivity of the created Si_Ag$_2$S devices and indicates large photoresponse. b) A schematic representation of the cross section of a Si_Ag$_2$S device (upper panel), here the Ag$_2$S QDs schematically presented as sequence of red spheres on top of Si. The intermediate and bottom panels present an energy diagram (ED) sketch of the Si_Ag$_2$S structure under direct and reverse current bias respectively. In the vicinity of the Ti/Si junction a space charge region is formed due to the redistribution of electrons between Si and Ti. This fact is reflected on ED by the deformation of E$_c$ and E$_v$ band levels.

The sketch of the energy diagram presented in **Figure 3**b describes the behavior of the IV curve of the Si_Ag$_2$S device in the case of positive and negative bias. When a positive potential is applied to the Ti/Au contact the IV curve displays a pronounced exponential character, as predicted by the thermionic emission model. The deviation from a truly exponential shape is caused by the variance of the Schottky contact area, which decreases in response to the increase of the forward bias voltage. Indeed, the contact area is proportional to the transfer length $l_t = (\rho_c(V_j)/R_{sq})^{0.5}$ defined by the sheet resistance of the Si wafer ($R_{sq}$) and the voltage-dependent contact resistivity of the Schottky contact ($\rho_c$).[35] Moreover, the latter quantity depends on the voltage applied to the Ti/Si junction as $\rho_c(V_j) = \eta V_t/J_s \cdot \exp\left(-V_j/(\eta V_t)\right)$, where $\eta$ is the ideality factor, $V_t$ is the thermal voltage and $J_s$ is the saturation current density.[36] Given that the Si_Ag$_2$S device is essentially presented by two face-to-face connected Schottky diodes, an increase of the built-in potential ($\varphi_{bi}$) at the Si/Ti interface, at one end of the junction with a decrease of the $\varphi_{bi}$ at the other end is induced, when a negative potential is applied to it. In the case of small bias voltages, the IV curve is almost linear when a negative potential is applied to the Si/Ti interface. This behavior of the IV curve suggests the hole conductivity of the created Si_Ag$_2$S devices. Thus the presence of a high response of the Si_Ag$_2$S device above 1.25 μm is explained by formation of impurity states in the band-gap of Si due to Ag$_2$S QDs surface states. When an IR photon is absorbed by an electron in the Si valence band, the generated carrier transits at impurity state Ag$_2$S in band-gap of the Si, this process owing to high carrier mobility in Si affects the built-in potential φ$_{bi}$ near the Ti/Au contacts.[37] The existence of the built-in potential φ$_{bi}$ near the Ti/Au contacts explains the zero bias mode of the devices.

Because of their non-stoichiometry, the Ag$_2$S QDs have a high concentration of trap states. These trap states along with surface states in Si also could contribute to the Si_Ag$_2$S detection above 1.1 μm. Since quantum dots do not come into contact with each other, the charge carriers from the Ag$_2$S trap states generated by the photoelectric effect and entering the Si increase its conductivity. The time constant of the Si_Ag$_2$S detector will be determined by the characteristic lifetimes of the free charge carriers in the silicon/metal regions.

The noise equivalent power (*NEP*) of the Si and Si_Ag$_2$S devices was measured at 1.55 μm in order of device quantity evaluation.

$$NEP = \frac{P}{V_s} \frac{V_n}{\sqrt{\Delta f}}, \frac{W}{\sqrt{Hz}}$$

A thermally stabilized laser diode at 1.55μm was used as a source. During measurements of the voltage responsivity, the laser diode power *P* was chosen so that the signal to noise ratio $V_s/V_n$ did not exceed 1.5–2. This guaranteed the linear operation mode of the studied devices. The signal $V_s$ and noise $V_n$ voltages were measured at a modulation frequency of the IR radiation *f* of 80 Hz. To detect the response, we used a selective voltmeter with an input resistance of 20 MOhm as a measuring instrument. The Si_Ag$_2$S detector had a resistance of 7MOhm at 300K. During the measurements, the Si_Ag$_2$S detector was at zero offset. The voltmeter was directly connected to the Si_Ag$_2$S detector. A signal with a voltage of 6.9 mV was recorded at the input of the selective voltmeter with a test radiation power of 5 μW at the input. In the absence of test radiation, the voltage at the input of the selective voltmeter (the detector noise) was of about 1 μV, hence the response in A/W units can be calculated as 70 μA/W with a noise level of 5*10^-14 A. The experimentally measured *NEP* for Si and Si_Ag$_2$S structures were 2.1 * 10$^{-8}$ W/√Hz and 4.5 * 10$^{-10}$

W/√Hz, respectively. Given the Si_Ag$_2$S detector had a resistance of 7MOhm at a temperature of 300 the Johnson noise value for such a Si_Ag$_2$S detector is 1.8 * 10 ^ -8 V. The measured value was about 1 10^-6 V at a frequency of 80 Hz. Assuming that the spectral noise of the Si_Ag2S detector follows a 1/f law, the lowest noise level can be achieved already in the KHz frequency range of the test radiation modulation for such a Si_Ag$_2$S detector. Thus estimated NEP for the Si_Ag2S detector becomes of the order of 10-13 W /√Hz. The specific sensitivity of the Si_Ag$_2$S device is of the order of $10^{11}$ cm√HzW$^{-1}$, which is comparable with values of the best commercially available IR detectors.

In conclusion, this work presented a simple, low-cost and effective technology of extending the photoresponse of Si towards the IR range. The technology based on Ag$_2$S quantum dots allows creating controllable surface states in Si. At 1.45 μm the response of the the Si_Ag$_2$S detector exceeds that of a conventional Si detector by a factor of 40. The specific sensitivity of the Si_Ag$_2$S device is of the order of $10^{11}$ cm√HzW$^{-1}$ at 1.55μm, which is comparable with values to the best commercially available IR detectors. Further development of this work towards the enhancement of the device sensitivity is concerned with optimizing the density of surface states in Si by using a mixture of quantum dots made from materials with different photoelectric thresholds of the internal photoelectric effect for the multi-color detection mode.

**Experimental Section**

The aqueous synthesis of colloidal Ag$_2$S QDs was based on the initial reagents of AgNO$_3$ and Na$_2$S. In the first case, 0.262 g AgNO$_3$ and 0.276 g TGA dissolved in 50 ml of water were poured into a thermostatically controlled reactor at 25 °C containing 200 ml of water. Next, the pH in the reactor with dropping of 1M NaOH solution was adjusted to 10. After that, using a peristaltic pump, 50 ml of a solution containing 0.18 g of Na$_2$S was infused into the reactor for 240 s. The solution in the reactor changed color from pale yellow to dark brown. To remove the reaction by-products, acetone was added to the 50% aqueous colloidal Ag$_2$S QDs solution and centrifuged. In this case, QDs settled to the bottom of the tube, and the solution with undesirable components was removed. The resulting Ag$_2$S QDs were redissolved in distilled water. The cleaning procedure was repeated several times.

The optical absorption spectra of the fabricated Ag$_2$S QDs were investigated with a USB2000+ spectrophotometer with USB-DT emission source (Ocean Optics, USA). Photoluminescence spectra were obtained on the automated spectral complex based on the diffraction monochromator MDR-4. The highly stable low-noise photodiode PDF10C/M (ThorlabsInc., USA) with a built-in amplifier was used as a photodetector for the near-IR region. The laser diode LPC-836 (Mitsubishi, Japan) operating at 660 nm was used as a source of photoluminescence excitation.

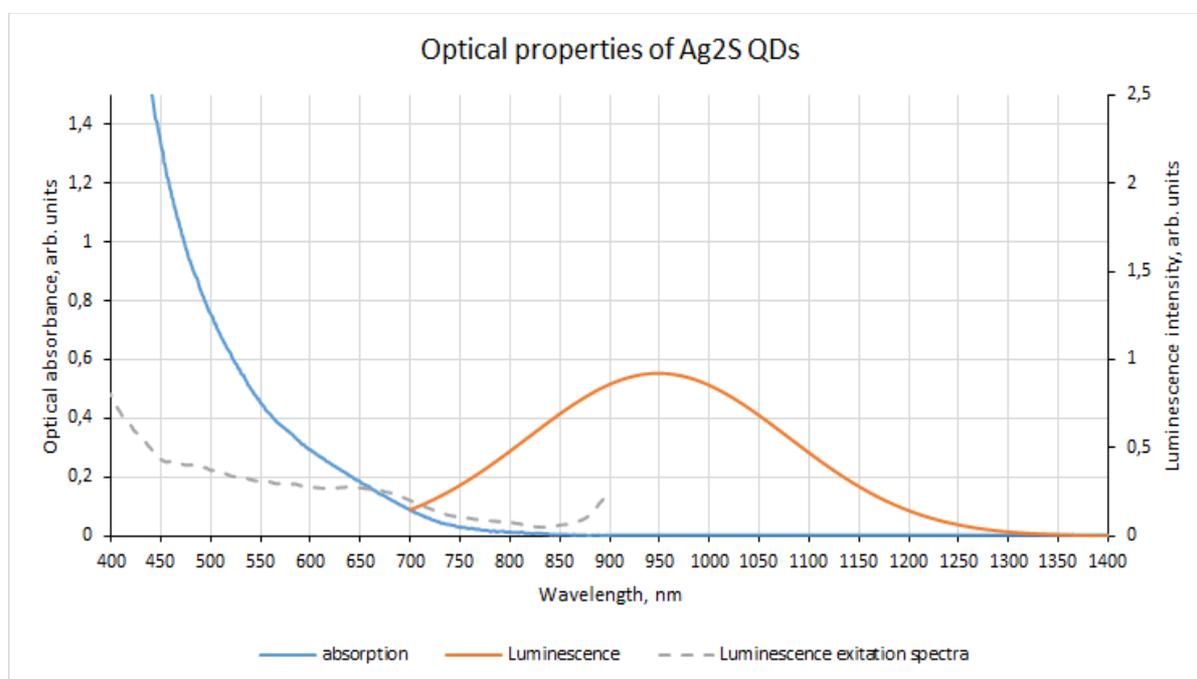

**Figure 4.** The optical properties of Ag2S colloidal solution of QDs.

The position of the characteristic feature in optical absorbance is 1.88 eV. which exceeds the band gap of an Ag2S single crystal with a monoclinic crystal lattice (1.0 eV). The observed difference is caused by the size effect. Estimates of the size of QDs, made using the Kayanuma formula, showed that particles with an average size of 2.4 nm correspond to the occurrence of these features, which is consistent with the results of SEM image analysis. The presence of a non-zero optical density at a wavelength of more than 700 nm can be due to both the dispersion of QD in size and impurity absorption of light. In the luminescence spectra of the Ag2S QDs samples under study, a wide band with a maximum at 950 nm was observed excited by radiation with a wavelength of 660 nm. A significant Stokes shift relative to the position of the exciton transition in the optical absorption spectrum and the half-width of the luminescence spectrum indicate the recombination nature of the observed emission. Studies of the photoluminescence excitation spectrum (dashed curve in Figure 4) show the existence of a number of features pertaining to both the intrinsic light absorption region in the 400-700 nm range and the impurity absorption region. The observed growth in the long-wavelength region of the spectrum (700–900 nm) is due to the possibility of direct excitation of the luminescence centers. Thus, the possibility of direct excitation of recombination photoluminescence by light quanta with a wavelength in the region of impurity absorption of light has been established.

The Si surface cleaning process was performed in Corial 200R, sequentially ion etching in $O_2$ for 15 s at 50 W and ion etching in Ar for 20 s at 50 W. Liquid etching with hydrofluoric acid ($HF:H_2O=1:10$) took 30 s. The contacts to the Si structure were formed by thermal deposition of Ti and Au to thicknesses of 5 nm and 200 nm respectively. $Ag_2S$ QDs were deposited on top of the cleaned surface of of a Si wafer by centrifuging for 1 min at 2000 rpm and heating to 120 °C on hotplate.

Structural properties of the centrifuged $Ag_2S$ QDs were studied with the standard techniques of TEM and XRD analysis. The sizes of the synthesized ensembles of colloidal $Ag_2S$ QDs were determined with the TEM Carl Zeiss Libra 120 operating at 120 kV. High resolution TEM images were obtained with a JEOL HRes Carl Zeiss Libra microscope 200 kV. The crystal structure was investigated with the X-ray diffractometer ARL X'TRA for $K_{\alpha 1}$ of copper.

Optical spectral and NEP measurements were performed with the infrared spectrometer with a Xe lamp as a IR radiation source and the laser diode TeraXion PS-LM-1550.12-40-06 (from PureSpectrum). Precise attenuation and measurement of the IR power was done with the tunable attenuator EXFO FVA-600 and OPHIR VEGA ROHS respectively. A hyper-hemispherical Si lens ($\rho$> 10 KΩ*cm) with a diameter of 12 mm and an extension for wafer with a thickness of 350 μm was used for the coupling the device with IR radiation. The IV curves of the device were measured by Keithley 2410. The signal and noise voltage within spectral and NEP measurements were measured by the Lock-in Amplifiers SR830 and UNIPAN 233 .

Acknowledgements

The study of the spectral response of the devices was made with the support of the Russian Science Foundation (project No. 17-72-30036), the study of the NEP and the responsivity was carried out with the support of the Ministry of Education and Science of the Russian Federation [State Task number 11.2423.2017/4.6]. The device design and fabrication were made with the support by Russian Science Foundation project No.16-12-00045.